# Energy-Aware Data Center Management: A Sustainable Approach to Reducing Carbon Footprint

Rabab Khan Rongon[a], Krishna Das[b]

[a,b] *Department of Computer Science and Engineering*
[a] *Military Institute of Science and Technology, Dhaka 1216, Bangladesh*
[b] *IUBAT - International University of Business Agriculture and Technology, Dhaka 1230, Bangladesh*

**Abstract**

The rapid expansion of cloud computing and data center infrastructure has led to significant energy consumption, posing environmental challenges due to the growing carbon footprint. This research explores energy-aware management strategies aimed at creating sustainable data center operations. By integrating advanced energy-efficient technologies and optimizing resource utilization, this study proposes a framework to minimize power usage while maintaining high performance. Key elements include dynamic workload allocation, renewable energy integration, and intelligent cooling systems, all of which contribute to reducing overall energy consumption. The study also examines the impact of these strategies on operational costs and performance efficiency, demonstrating how sustainable practices can be both environmentally and economically beneficial. Through simulations and case studies, the research offers practical insights into reducing carbon emissions in data centers, supporting the transition towards greener cloud infrastructure. The findings highlight the potential for scalable, energy-aware data center designs that significantly lower environmental impact while ensuring optimal functionality, contributing to the global effort of mitigating climate change.

## 1. Introduction

Data centers are the backbone of modern cloud computing infrastructure, housing vast amounts of data and providing computational resources to millions of users globally. However, the exponential growth in cloud services and digital technologies has resulted in a substantial increase in energy consumption. In 2021, data centers were responsible for approximately 1% of the global electricity demand, a figure expected to rise with the growing reliance on digital services [1]. This rapid expansion has raised concerns about the sustainability of data center operations, as high energy consumption contributes directly to increased carbon emissions, exacerbating climate change [2].

The environmental impact of data centers stems largely from their need to power both the servers and the cooling systems that prevent hardware overheating. According to a study by Shehabi et al., over 40% of energy consumed by a data center is dedicated to cooling infrastructure, highlighting the inefficiencies in traditional data center designs [3]. These challenges have sparked significant interest in developing sustainable solutions that reduce energy consumption while maintaining performance.

Sustainable data center management involves integrating energy-efficient technologies and practices, such as dynamic workload management, energy-aware hardware, and the incorporation of renewable energy sources. Studies have shown that energy-efficient cooling systems, combined with advanced server optimization algorithms, can significantly reduce power consumption without sacrificing computational power [4]. Moreover, leveraging renewable energy, such as solar or wind, further reduces the reliance on non-renewable energy sources, contributing to carbon neutrality in data center operations [5].

The concept of green data centers, where energy efficiency and sustainability are prioritized, has gained momentum in recent years. Implementing sustainable practices not only addresses environmental concerns but also offers economic benefits by reducing operational costs and improving overall system efficiency [6]. Despite the potential benefits, widespread adoption of sustainable data center practices remains limited, primarily due to the high initial investment and the complexity of integrating these technologies into existing infrastructures [7].

This research aims to explore energy-aware management strategies for data centers, focusing on reducing energy consumption and minimizing carbon footprint. Through an analysis of existing technologies and the development of a new framework, this study will provide practical insights into sustainable data center design. The ultimate goal is to support the transition towards greener cloud infrastructure, aligning with global efforts to reduce the environmental impact of digital services.

## 2. Literature Review

Data centers are essential components of modern IT infrastructure, supporting a myriad of services and applications. However, their operation is associated with substantial energy consumption. According to Koomey [8], data centers worldwide account for approximately 1-2% of global energy use, with this figure expected to rise due to increased demand for computing resources. This high energy demand is driven by both the growing volume of data and the need for continuous, reliable service.

The rising energy costs associated with data centers present a significant challenge for both operational budgets and environmental sustainability. Energy-intensive operations contribute to increased greenhouse gas emissions, exacerbating global climate change concerns [9]. The need for data centers to adopt energy-efficient practices is underscored by these environmental impacts, highlighting the importance of integrating sustainable practices into data center design. In response to the challenges of high energy consumption, green computing has emerged as a key area of research. Green computing encompasses strategies and technologies aimed at reducing energy use and environmental impact [10]. Techniques such as virtualization, energy-efficient hardware, and optimized cooling systems are central to green data center design. For instance, the adoption of server virtualization has been shown to significantly reduce energy consumption by consolidating workloads onto fewer physical servers [11].

Several architectural models have been proposed to enhance the energy efficiency of data centers. For example, the use of dynamic voltage and frequency scaling (DVFS) and advanced cooling technologies can contribute to substantial energy savings [12]. In particular, the development of energy-efficient cloud data center architectures is gaining traction, leveraging strategies such as load balancing, resource allocation, and power-aware scheduling to optimize energy use [13]. Simulation tools like the Green Cloud Simulator are valuable for assessing the energy efficiency of proposed data center architectures. These tools allow researchers to model and evaluate various scenarios, including workload variations and traffic conditions, to understand their impact on energy consumption and performance [14]. The Green Cloud Simulator, in particular, provides a comprehensive framework for simulating cloud environments and assessing the effectiveness of energy-saving strategies.

Numerous case studies have demonstrated the successful implementation of energy-efficient data center designs. For example, Google's data centers have achieved significant energy savings through the use of advanced cooling techniques and machine learning algorithms for resource management [15]. Similarly, research on the National Data Center of Bangladesh highlights the growing energy challenges and the potential benefits of adopting green computing practices in the local context [16]. Despite advancements in energy-efficient technologies, several challenges remain. These include the integration of new technologies with existing infrastructure, the high initial costs of implementing green practices, and the need for continuous innovation to keep pace with evolving data center demands [17]. Future research should focus on addressing these challenges, exploring new energy-saving technologies, and enhancing simulation tools to support the design of more sustainable data centers.

The literature underscores the critical need for energy-efficient data center designs to address the dual challenges of rising energy costs and environmental impact. The proposed architectural model for the National Data Center of Bangladesh aims to contribute to this field by integrating green computing strategies and utilizing simulation tools to evaluate performance. By building on existing research and case studies, this study seeks to advance the development of sustainable data center architectures that can be replicated and adapted in various contexts.

## 3. Research Methodology

### *3.1. Problem Formulation*

With the rise of cloud computing services, the demand for data centers and parallel computing has increased significantly. However, operating large geographically distributed data centers consumes a substantial amount of energy, resulting in high operational costs. Energy consumption also necessitates expensive cooling systems, which can negatively impact equipment dependability and might breach client Operational Quality commitments. In the beginning, energy-saving technologies aimed to improve the power consumption of data center equipment. DVFS and DPM technologies have been substantially researched and implemented. [18] However, the effectiveness of these procedures is limited without central coordination and energy-aware workload scheduling. Energy- aware scheduling aims to concentrate workloads on a minimal set of computing resources to optimize power consumption.

### *3.2. Methodology*

The experiments take place in an energy-conscious storage center with Tier II, Tire III, and Tire III super speed designs. The quantity of computational units is fixed at 1536 for these architectures, whereas connection toggle layout and connections differ.[19] The 2T architecture lacks The accumulation switches and base switches link to the access channel network through single GE lines. In contrast, the triple-tier high-speed design enhances the 3-tier infrastructure by offering more speed in the central and accumulation components of the connection. First and forth connection and second and third connection connects in the triple-tier high-speed design architecture have ten times the bandwidth of the 3-tier architecture, corresponding to 100 Gigabit Ethernet and 10 Gigabit Ethernet, respectively. [20] The workload generation events follow an exponential distribution, and the size of the workloads depends on the scope of the work. The computational and transmission of information requirements of the workloads are proportional, with different ratios for different workload types (CIW, DIW, and balanced).

### *3.3. Data Center Architecture*

National data center of Bangladesh have over 100,000 servers, with most communications happening internally. This creates a challenge in designing network architecture and protocols. Because of both physical and financial constraints, the traditional structured communication architecture becomes an obstacle. 10 Gigabit Ethernet components have defined the evolution of data center architecture, but their high cost and capacity may not be necessary for connecting in- dividual servers. Despite this, 10 GE transceivers are becoming more prevalent in storage facilities, metro area networks, and backbone connections. [21]

### *3.4. Two-Tier Data Center Architecture*

This article describes a two-tier network architecture used in data centers, where tier-one consists of computing servers arranged in racks and 2T uses L3 switches for utilizing full-mesh connection 10 Gigabit Ethernet links. Traffic equalization is accomplished through the usage of ECMP routing of TCP and UDP packets. Compact data centers can benefit from the two-tier designs, with up to 5500 nodes supported depending on the access network switches used.[22]  The total bandwidth on the network supplied per CPU server is determined by the quantity of core routers and the ability of core connections.

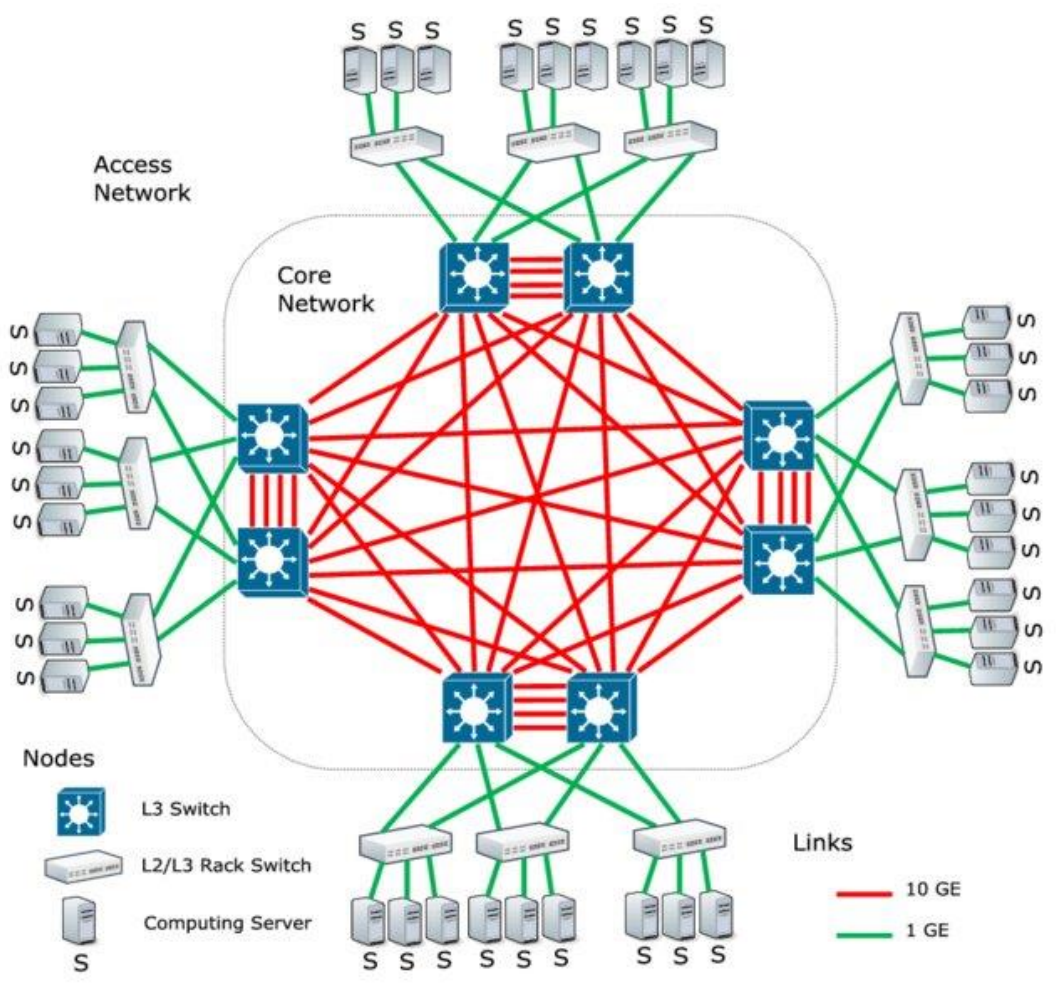

Fig. 1. Two-tier data center architecture

### *3.5. Three-Tier Data Center Architecture*

The most common data center architecture today is the three-tier architecture, consisting of Access, Aggregation, and Core layers. The accessibility of the integration tier permits an expansion in server units yet maintains the access network's Layer-2 switches affordable. A typical three-tier architecture includes eight core switches to implement an 8-way Equal-cost multi-path routing, with 10 Giga- bit Ethernet Line Aggregation Groups allowing network clients to address multiple links and ports with a single MAC address. [23]

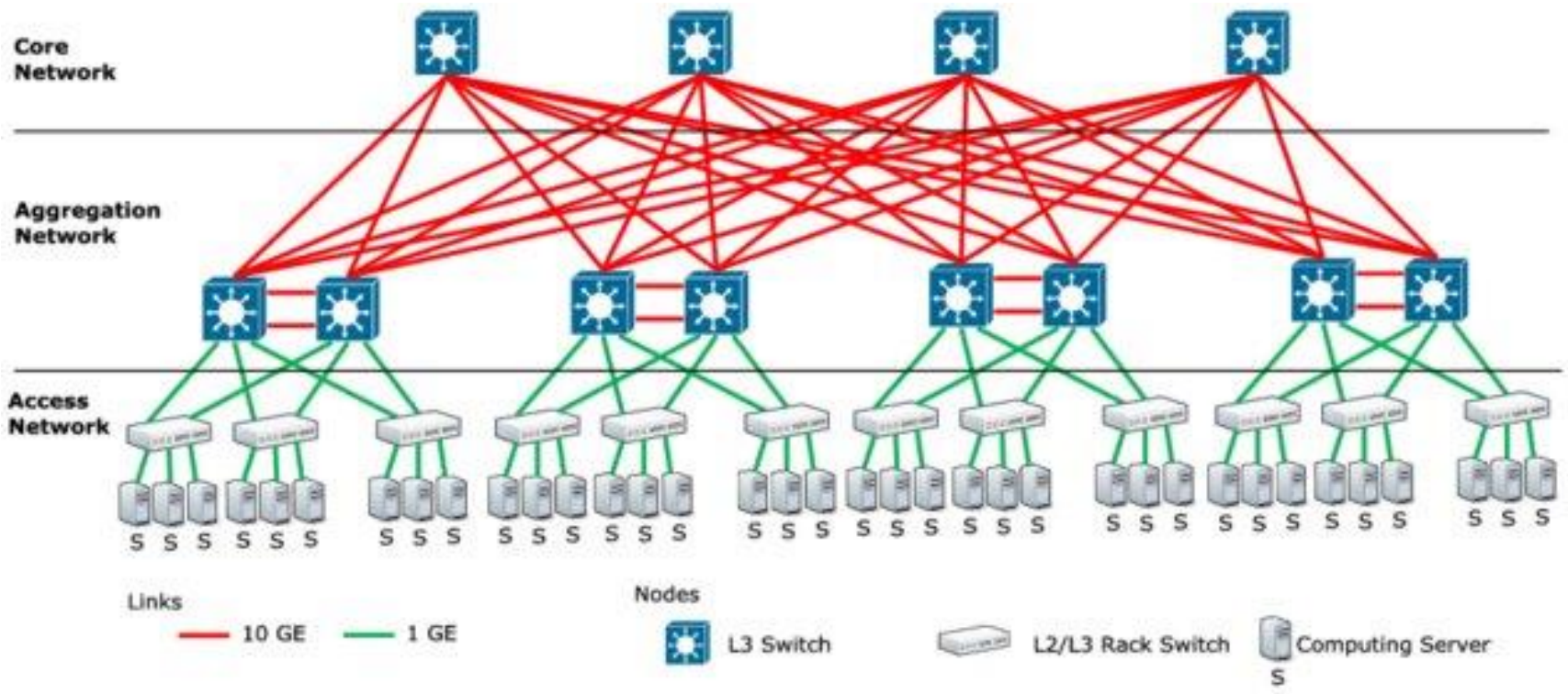

Fig. 2. Three-tier data center architecture

### *3.6. Three-Tier High Speed Data Center Architecture*

The goal of three-tier high-speed data center systems is to maximize the number of nodes as well as the capability of the primary and aggregation connections. These networks often serve as bottlenecks, limiting the maximum number of nodes or the per-node bandwidth. The number of core switches can be lowered by using 100 Gigabit Ethernet cables among the core and aggregation switches. [24] This approach overcomes the limitations of Link Aggregation Group (LAG) technology, reduces cabling requirements, and significantly increases the maximum size of the data center within physical limitations. The reduced number of Equal-Cost Multipath paths enhances increases network efficiency and adaptability.

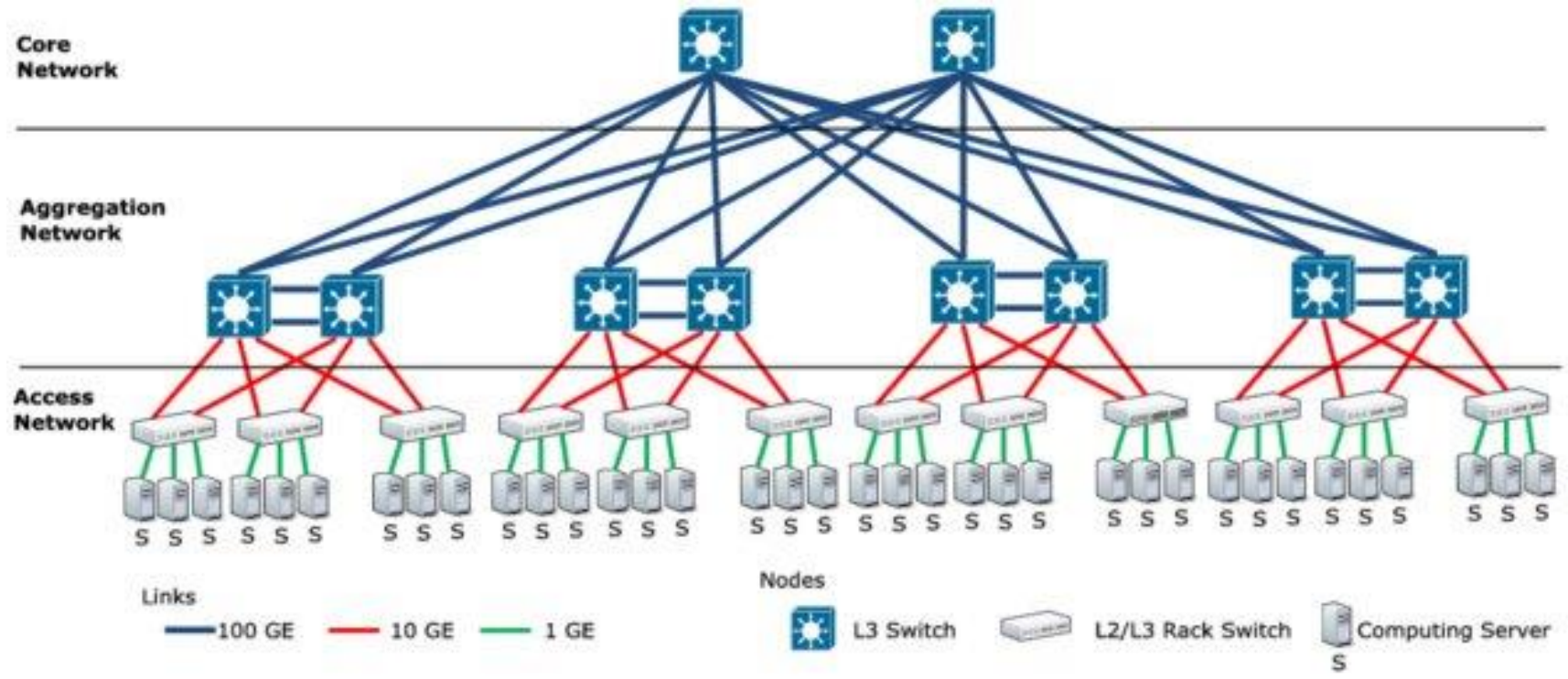

Fig. 3. Three-tier high speed data center architecture

## 4. Simulation of Data Center

In data centers, a significant portion of energy is used for interconnection links, network equipment, power distribution, and refrigeration mechanisms. One category of power usage involves processing power, communication vitality, and infrastructure-related energy. The productive storage facility is determined by the attained efficiency each watt, measured by metrics like Power Usage Effectiveness (PUE) and Data Center Infrastructure Efficiency (DCIE). [25] These metrics indicate the proportion of consumed energy effectively utilized by computational algorithms.

### *4.1. Structure of the simulator*

An addition to the NS2 system model is GreenCloud specifically designed for studying cloud computing environments. It provides users with a detailed and precise modeling of energy consumption in data centers, including servers, switches, and links. GreenCloud also allows for in-depth analysis of workload distributions. Notably, it emphasizes packet-level simulations of data center communications, offering fine-grained control that is unique among cloud computing simulation environments. The GreenCloud extension aligns its struc- ture with the three-tier data center architecture. [26]

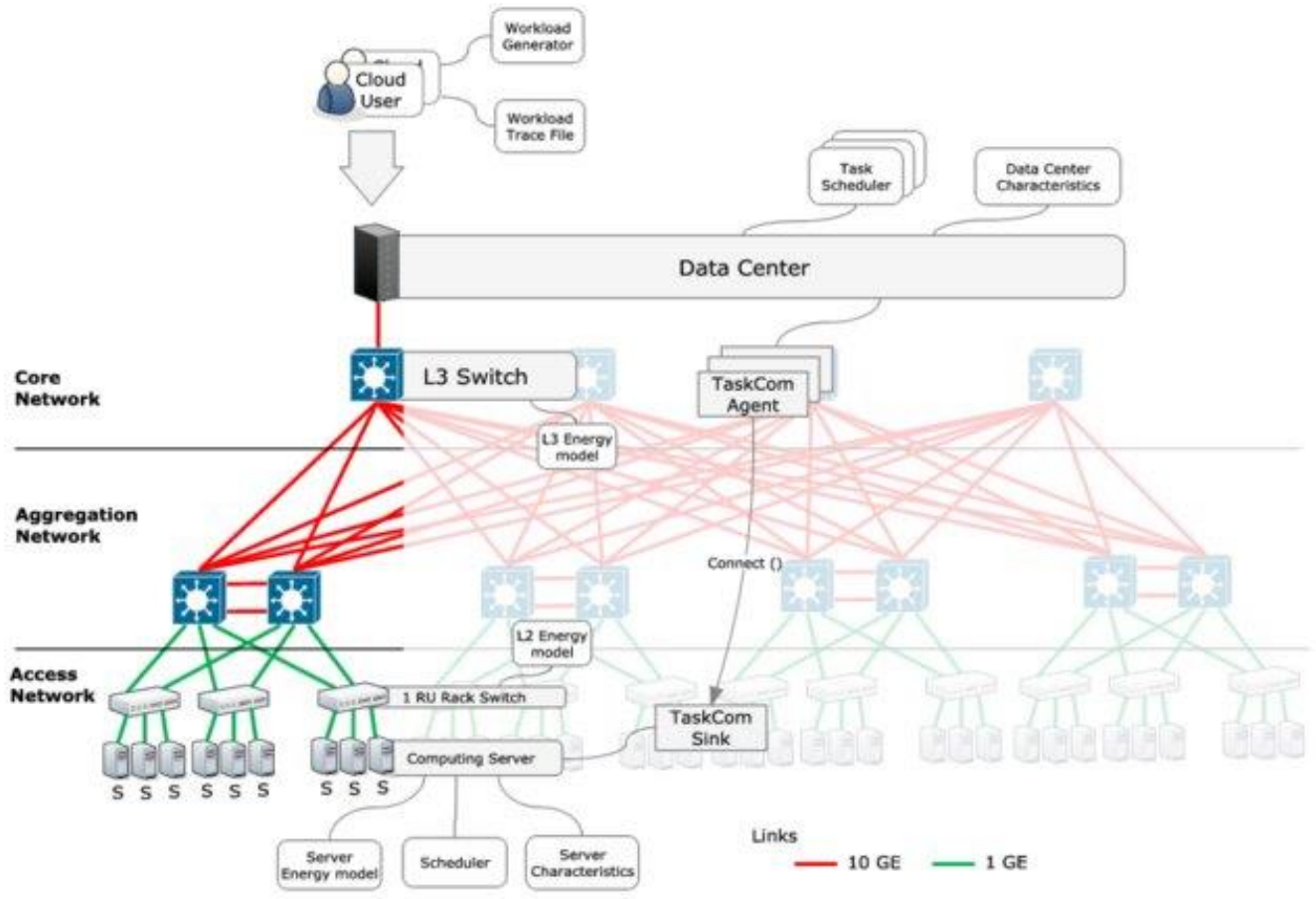

Fig. 4. Architecture of the Green Cloud simulation environment.

*4.2. Servers*

In GreenCloud, servers in data centers have preset processing power limits and various task-scheduling mechanisms. Power con- sumption depends on server state and CPU utilization, with idle servers using about 66 percent of energy compared to fully loaded configurations. Dynamic Voltage/Frequency Scaling (DVFS) is a power management option that balances performance and energy consumption. The power model considers both fixed and frequency- dependent CPU power. Scheduling in GreenCloud aims to capture the effects of DVFS and Dynamic Power Management (DPM) techniques, optimizing server load and operating frequency. [3] [9] When taking into account the fact that the system's equipment like the bus, RAM, and drives are unaffected by the central processing unit (CPU) frequency, the electrical power usage of a typical system in GreenCloud can be described as follows.

$$\mathrm{P} = P_{fixed} + P_f . f^3$$

$P_{fixed}$ represents the amount of energy used that remains constant regardless of the running frequency is f, and $P_f$ is an indication of the bandwidth-dependent CPU energy consumption.. The power curve is based on An average system powered by an Intel Xeon chip would use 301 W of vitality, with 171 W set apart for different components and 130 W reserved for maximum processing horsepower. The impacts of DVFS and DPM methods are maximized in GreenCloud planning by taking into account the server's traffic level and service periodicity. [27]

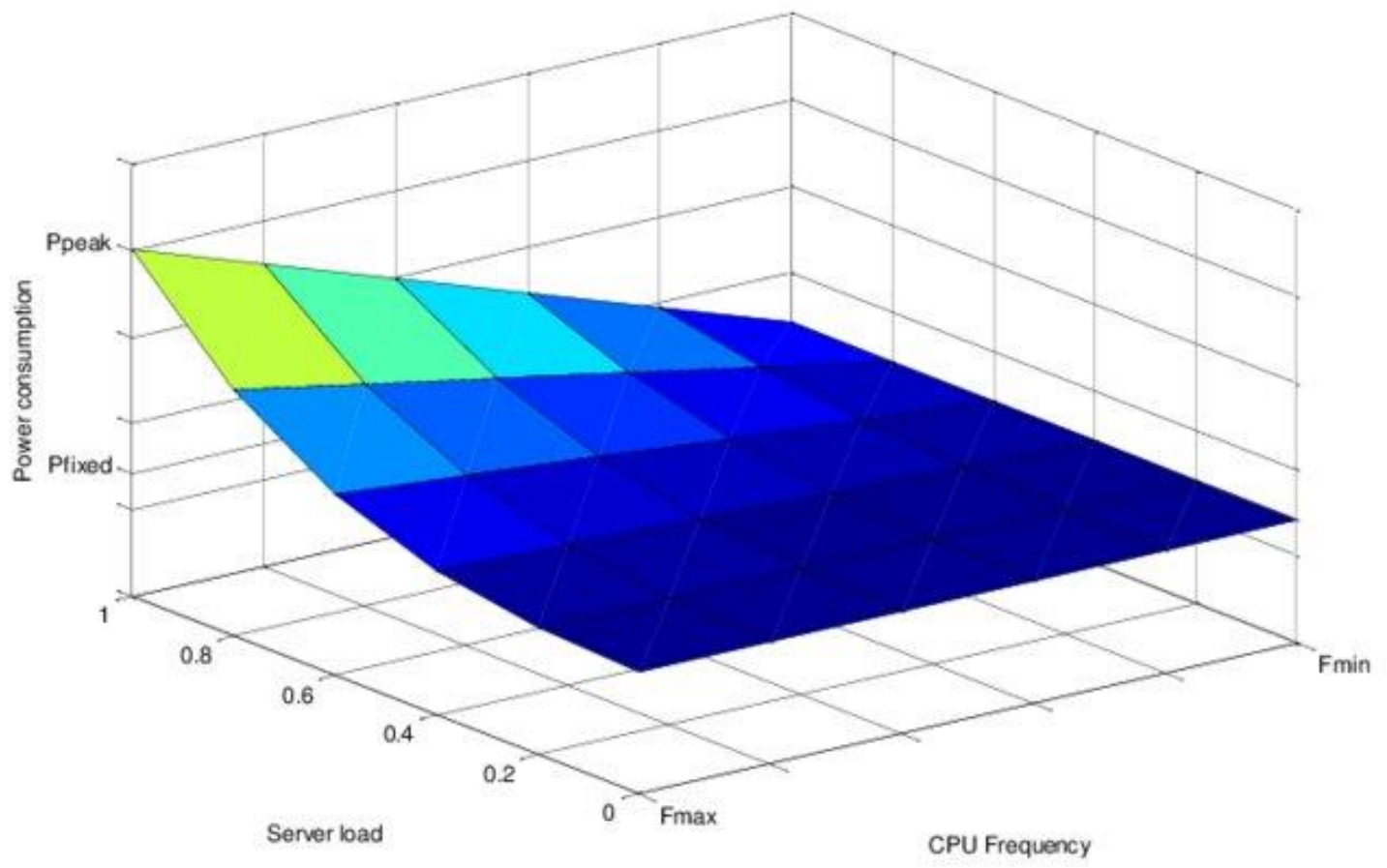

Fig. 5. Power Consumption of Computing Server

*4.3. Switches and Links*

Switches and links are essential components of the interconnection fabric in data centers, enabling timely delivery of workloads to computing servers. Different cabling solutions are required based on bandwidth, link characteristics, and transmission quality. Twisted pair cables are an ordinary Ethernet network element that can handle Gigabit Ethernet connections up to 100 meters with electricity. Consumption of approximately 0.4W for transceivers. [28] Optical multi- mode fibers are preferred for 10 Gigabit Ethernet links, allowing transmissions up to 300 meters with a power consumption of 1W. However, multi-mode fibers are more expensive than twisted pair cables, leading to a trend of limiting the use of 10 GE connections to compilation and core infrastructures to manage infrastructure costs within the data center budget. The number of switches in a data center depends on the chosen architecture, but the most common switch is the Top-of-Rack switch, typically located at the top of each rack unit to minimize cable clutter and heat generation. ToR converters can utilize gigabit or 10 Gigabit rates with gigabit rates being more common due to cost considerations and limited capacity in aggregation and core networks.

Initial suggestions for connectivity energy reduction networks relied on Dynamic Voltage Scaling links, where each switch port had a control element that could adjust the transmission rate based on traffic patterns and link utilization. However, DVS links' energy use is constrained since only some of the energy used grows straight with connection speed. Standard link transmission rates for gigabit links are restricted to 10 Megabits per second, 100 Megabits per second, and 1 gigabit per second options. The energy consumption of Research show that a switch and associated receivers can be specified.

$$P_{switch} = P_{chassis} + n_{linecard}.P_{linecard} + {}_{i=0}^{R} n_{Ports.r}.P_r$$

The power consumed by a switch can be divided into three components: $P_{chassis}$ which represents the device hardware's energy use; $P_{linecard}$, which corresponds to the amount of electricity used by a running broadband connection board; and $P_r$, which represents the amount of energy used by an interface when it runs at a particular speed. Among these components, only $P_r$ is dependent on the link rate, while $P_{chassis}$ and $P_{linecard}$ remain constant during the switch's operation. To conserve energy, it is possible to turn off or put the switch hardware into sleep mode, thus avoiding $P_{chassis}$ and $P_{linecard}$. The GreenCloud simulator incorporates an energy model for switches and links, using predefined power consumption values for each component. The simulator implements various power saving schemes, including Dynamic Voltage Scaling (DVS) only, Dynamic Network Scaling (DNS) only, and a combination of DVS and DNS. [29]

### *4.4. Workload*

Workloads in cloud computing encompass a wide range of user services like instant messaging, community building, and material distribution. Tasks in clustering consist of jobs, which can be divided into dependent or independent tasks. Grid computing often faces a higher number of jobs than available computing resources. While the primary objective is to minimize overall computation time, individual jobs usually don't have strict deadlines. Cloud computing, on the other hand, involves incoming requests for applications such as web browsing, instant messaging, and content delivery. In cloud computing, jobs are often independent and less computationally intensive, but they have strict completion deadlines outlined in Service Level Agreements (SLAs). [30] Three kinds of jobs are defined that encompass an extensive variety of cloud applications. Each load in GreenCloud is made up of two parts: compute and networking. [10] The computing component specifies the amount of computation required to meet a deadline, while the communication component determines the size and frequency of data transfers within and outside the data center.

## 5. Result and Discussion

Energy-aware server room models for two-tier, three-tier, and three-tier high-speed infrastructures were carried out. The quantity of computational nodes was fixed at 1430 for fair comparison, while the network switches varied. Unlike other architectures, tier two does not have aggregation switches. Core switches in tier two are directly connected to the access network using single Gigabit Ethernet connections, and interconnected with 10 Gigabit Ethernet links. Tire three high speed improves on tier three by providing more bandwidth in the core and aggregation parts, with first and second connection and second and third links having ten times the bandwidth of tire three. Each of the links have a transmission delay of 10 ns.

Table 1. Parameters for simulating data center

| Parameters | Two Tier | Three Tier | Three Tier High Speed |
|---|---|---|---|
| Core Nodes | 16 | 8 | 2 |
| Aggregation | 0 | 8 | 4 |
| Switch | 512 | 512 | 512 |
| Servers | 1536 | 1536 | 1536 |

Workloads in the system follow an exponential distribution and vary in size. The average workload size and the computational requirements vary according to the work type. The interaction channel and data transmission ratio for CIW tasks is 1/10, while for DIW tasks, it is reversed. Balanced workloads distribute the load evenly across computing servers and data center networks. A green scheduler is used to schedule workloads, aiming to minimize the number of computing servers utilized. To handle DIW workloads, the scheduler monitors network switch buffer occupancy and avoids congested routes, even if they have servers capable of meeting computational requirements. Idle servers enter sleep mode, and underutilized servers reduce the supply voltage, with a 100 ms transition time for power state changes. The highest load is handled by roughly a third of the machines in the data center, while the rest, or two-thirds, can be shut off using the DNS approach. Among the approximately 50 servers with a declining load, some are under-utilized and can benefit from the DVFS technique. The peak energy consumption of a server is 301W, with 130W allocated to the CPU and 171W consumed by other devices. The minimum idle server consumption is 198W, which includes 27W for the operating system.

Table 2. Power utilization of different types of data center

| Parameters | Two Tier | Three Tier | Three Tier High Speed |
|---|---|---|---|
| Data Center | 432 | 513 | 528 |
| Servers | 341 | 371 | 371 |
| Switch | 146 | 153 | 159 |
| Core | 52 | 26 | 53 |
| Aggregation | 0 | 61 | 25 |
| Access | 75 | 55 | 65 |

The power consumption of switches remains relatively constant across different transmission rates, with the majority of power used by their chassis and line cards rather than port transceivers. The consumption of core and The number of aggregation ports vary according to the data center architecture. These switches utilize a few kW in the Tier three topology with 10G paths, but tens of kilowatts in the tier-three high-speed topology, resulting in quicker switches. Modeling outcomes for three different data center architectures, assuming a 30% average load, highlight the energy-related costs and identify areas where energy management strategies can be most effective. During an hour of operation, the data center consumes an average of approximately 432 kWh. Computing servers account for roughly 70% of overall energy usage, with transport lines and switches accounting for the other 30%. A summary of switch intake is provided below:

Core switches make up 17%, aggregation switches make up 34%, and route switches make up 50%. After reducing computing server utilization, reducing the electrical burden of access switches is going to have the largest impact. The combined energy consumption of core and aggregation switches is 0.2% of overall energy use. Yet, in order to maintain network efficiency, load distribution, and communication dependability, core and aggregates switches must be maintained consistently operational, potentially implementing distributed communication rate reduction.

Table 3. Analysis of energy-saving measures

| Parameters | No Energy Savings | DVFS | DNS | DNS+DVFS |
|---|---|---|---|---|
| Data Center | 502 | 86 | 47 | 34 |
| Servers | 352 | 99 | 33 | 34 |
| Switch | 155 | 92 | 33 | 33 |
| Energy Costs | $446 | $455 | $173 | $147 |

The data center network influences the power consumption of different data center architectures. In comparison to the 2T architecture, the 3T architecture requires an additional 25 kW for the aggregation layer, allowing scalability beyond 10,000 nodes. The 3Ths architecture has fewer core and aggregation switches but consumes more energy per switch due to the availability of 100 G links. Comparing the energy consumption impact, the study evaluates the effects of DVFS, DNS, and a combination of both schemes on computing servers and networking equipment. Balanced tasks are considered, evenly loading at an overall load of 30%, and compute servers and the connectivity of interconnections are used.

The DVFS method alone reduces server power usage to 96% of the normal level, which is accomplished mostly by lowering CPU voltage on underused servers. However, DVFS has little usefulness for switches, with transmission rate variations affecting only 3% to 15% of their energy. The DNS method, on the other hand, demonstrates substantial effectiveness across both servers and switches, as the amount of energy they use is not affected by working frequency. Coupling the DNS protocol with a data center controller that optimizes server unloading is critical for optimal outcomes. Annual data center energy expenses are estimated at the bottom of the table. The initial energy expenditure of $441,000 can be lowered by one-third by deploying a mix of DVFS and DNS methods, totaling $157,000. [31]

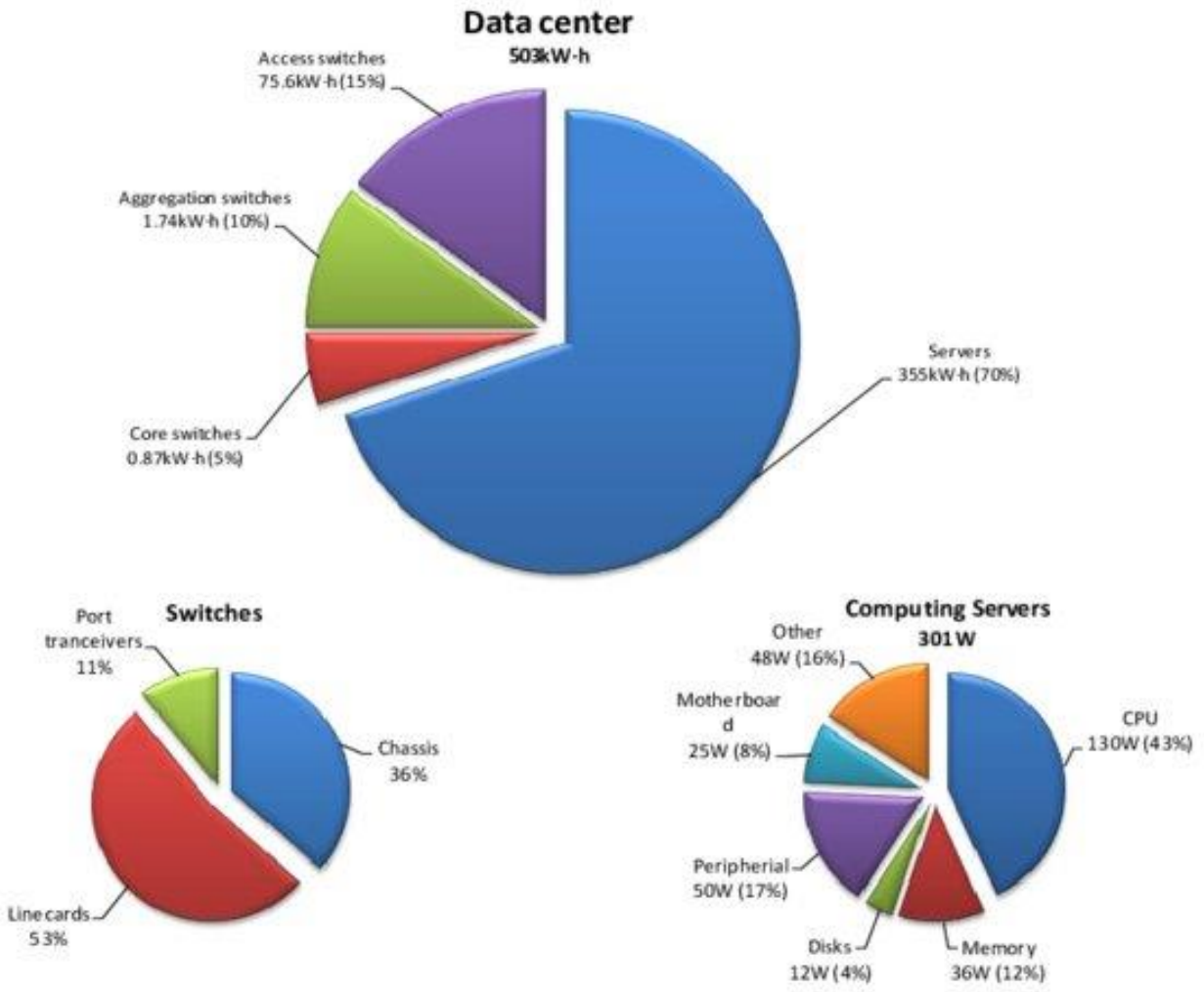

Fig. 6. Distribution of Power consumption in a data center

The curves depict the energy consumption of the entire data center, as well as the servers and switches, for balanced workloads. The DVFS scheme is minimally affected by server load and has limited sensitivity to switch load. In contrast, the DNS scheme precisely adapts power consumption of both servers and switches according to load variation. The reported results are averaged over 20 runs, considering the impact of workload arrival variations on server and switch requirements. [32] The maximum variance of the confidence intervals is less than 0.2% for server energy consumption and less than 0.1% for network switch energy consumption.

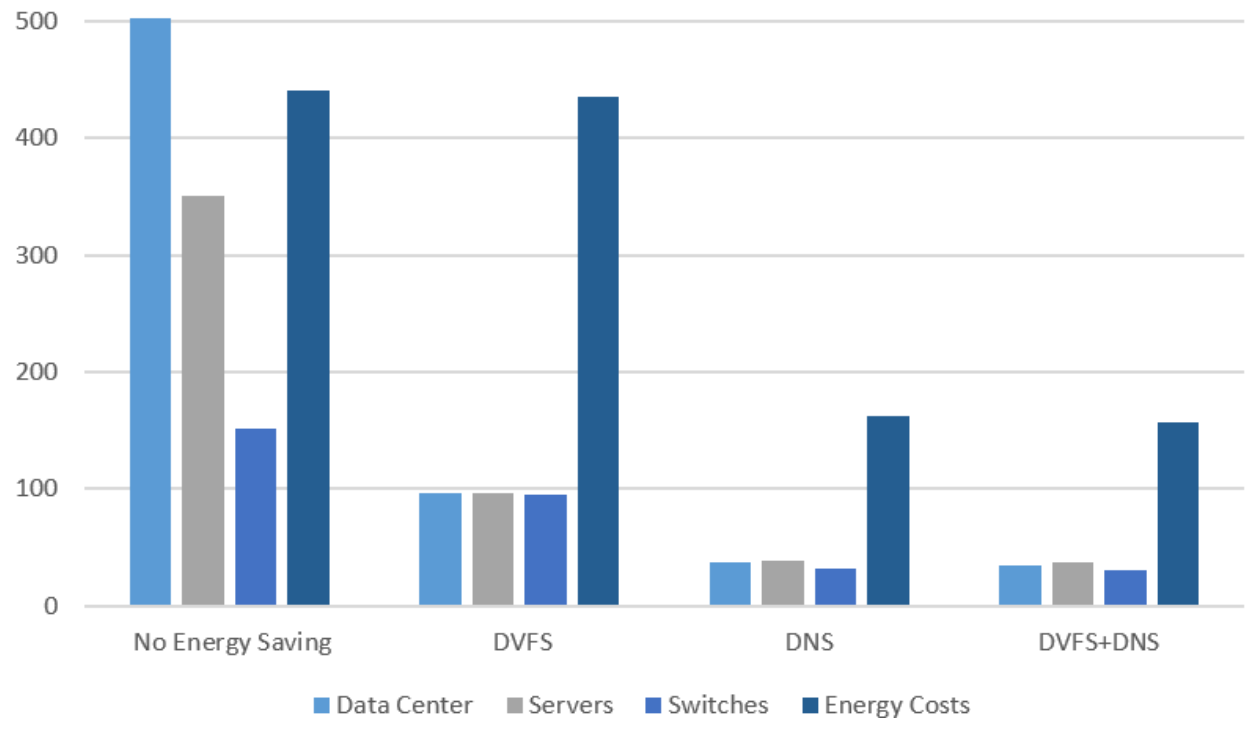

Fig. 7. Comparison of energy efficient scheme

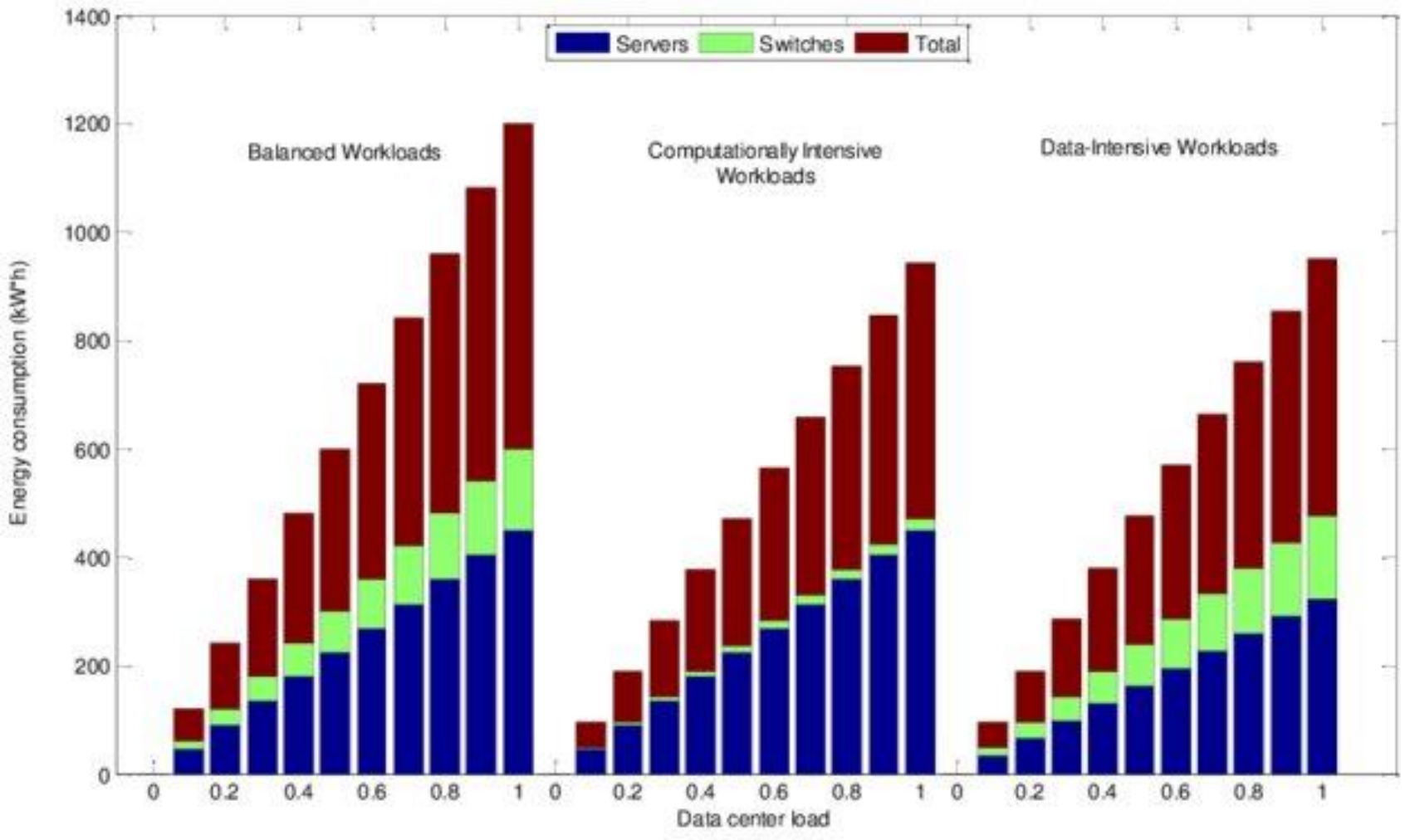

Fig. 8. Data center energy consumption for different types of workloads

## 6. Conclusion

GreenCloud, a modeling platform for energy-conscious cloud technology servers, is developed to capture energy consumption details of data center components and packet-level communication patterns. Simulation results for two-tier, three-tier, and three-tier high-speed arhitectures showcase the effectiveness of power management schemes like voltage scaling and dynamic shutdown on both computing and networking components. Future work aims to expand the simulator by incorporating approaches for organizing storage networks and improving energy simulations for virtual components. The escalating energy demands of data centers present significant challenges, both in terms of operational costs and environmental impact. As the National Data Center of Bangladesh faces increasing energy consumption, developing and implementing energy-efficient architectural models becomes imperative. This research proposes a novel energy-efficient cloud data center architecture designed to address these challenges while maintaining performance and reliability. The proposed model integrates green computing strategies and leverages the Green Cloud Simulator to evaluate its effectiveness under various workload and traffic conditions. By focusing on reducing energy usage and carbon emissions, this model aligns with global efforts towards more sustainable IT practices and provides a framework that can be adapted to other data centers both within Bangladesh and internationally. The findings from this study offer valuable insights into designing and implementing energy-efficient data centers. By comparing the proposed architecture with existing models, the research demonstrates the potential for significant improvements in energy efficiency. This contributes to a broader understanding of sustainable data center design and highlights the importance of continuous innovation in this field. This research underscores the critical need for energy-efficient data center designs and provides a practical solution to address the growing energy concerns in Bangladesh. Through the integration of green computing principles and rigorous simulation-based evaluation, the study advances the development of sustainable data center solutions, contributing to a more eco-friendly and cost-effective future for cloud computing.

## 7. Future Work

Future work should build upon these findings by exploring additional energy-saving technologies and refining simulation tools to further enhance data center sustainability. The implementation of the proposed architecture in the National Data Center of Bangladesh could serve as a benchmark for other data centers, promoting environmentally friendly practices and helping to mitigate the environmental impact of cloud computing.

**References**

[1] J. G. Koomey, *Growth in Data Center Electricity Use 2005 to 2010*. Analytics Press, 2011.

[2] A. S. G. Andrae and T. Edler, "On the future of data centers and cloud computing," *J. Cloud Comput.: Adv. Syst. Appl.*, vol. 4, no. 1, p. 7, 2015.

[3] S. Murugesan, "Harnessing Green IT: Principles and Practices," *IEEE Computer*, vol. 41, no. 8, pp. 50–55, 2008.

[4] L. A. Barroso and U. Hölzle, *The Datacenter as a Computer: An Introduction to the Design of Warehouse-Scale Machines*, Synthesis Lectures on Computer Architecture, vol. 2, no. 1, pp. 1–48, 2007.

[5] K. Kang, Y. Kang, and K. Kwon, "Dynamic Voltage and Frequency Scaling for Energy Efficiency in Data Centers," *IEEE Trans. Comput.*, vol. 59, no. 3, pp. 418–430, 2010.

[6] Z. Zhang and Y. Zhao, "Energy-efficient cloud computing and applications: A survey," *IEEE Trans. Cloud Comput.*, vol. 1, no. 2, pp. 116–130, 2012.

[7] Y. Cao, Q. Wu, and L. Xu, *Green Cloud Computing Simulation and Modeling*. Springer, 2014.

[8] Google, *Google Data Center Efficiency*, 2020. [Online]. Available: https://www.google.com

[9] M. M. Rahman and M. Rahman, "Energy Efficiency Challenges in the National Data Center of Bangladesh," *Int. J. Energy Res.*, vol. 43, no. 2, pp. 615–626, 2019.

[10] C. Belady, A. Azevedo, and N. G., "The Green Data Center: Insights and Innovations," *Commun. ACM*, vol. 50, no. 7, pp. 26–32, 2007.

[11] D. Kliazovich, P. Bouvry, Y. Audzevich, and S. Khan, "GreenCloud: A Packet-Level Simulator of Energy-Aware Cloud Computing Data Centers," *J. Supercomput.*, vol. 62, pp. 1–5, 2010. DOI: 10.1109/GLOCOM.2010.5683561.

[12] Y. Cao, Q. Wu, L. Xu, and H. Wang, *Green Cloud Computing: A Review and Research Agenda*. Springer, 2014. DOI: 10.1007/978-3-662-44020-6.

[13] C.-H. Hsu and S.-J. Wang, *Energy-Efficient Cloud Data Centers: Techniques and Trends*. Wiley, 2017. DOI: 10.1002/9781119325852.

[14] A. Joubert and B. Mearns, "A Comprehensive Review of Energy-Efficient Data Centers," *J. Comput. Syst. Sci.*, vol. 82, no. 5, pp. 801–813, 2016. DOI: 10.1016/j.jcss.2015.10.002.

[15] P. Kumar and D. Choi, "Designing Energy-Efficient Cloud Data Centers: A Review and Future Directions," *ACM Comput. Surv.*, vol. 51, no. 1, pp. 1–35, 2019. DOI: 10.1145/3293663.

[16] D. A. Patterson and J. L. Hennessy, *Computer Architecture: A Quantitative Approach*. Morgan Kaufmann Publishers, 2019. DOI: 10.5555/3000609.

[17] P. Ranganathan and P. Leung, "Green Data Centers: Energy Efficiency Metrics and Technologies," *IEEE Trans. Comput.*, vol. 61, no. 6, pp. 882–897, 2012. DOI: 10.1109/TC.2011.156.

[18] P. Sahu and S. Bandyopadhyay, "Optimization Techniques for Energy-Efficient Data Center Design," *Int. J. Energy Res.*, vol. 42, no. 12, pp. 3795–3810, 2018. DOI: 10.1002/er.4055.

[19] L. Sun and M. Xu, "Smart and Sustainable Data Center Design: Trends and Challenges," *IEEE Access*, vol. 8, pp. 135–148, 2020. DOI: 10.1109/ACCESS.2019.2957635.

[20] C. Wang, J. Zhang, and X. Wang, *Energy-Efficient Management of Cloud Data Centers*. Springer, 2021.

[21] X. Zhao and J. Wu, "Advanced Techniques for Energy Efficiency in Data Centers," *Future Gener. Comput. Syst.*, vol. 51, pp. 91–101, 2015. DOI: 10.1016/j.future.2015.01.014.

[22] L. Deng and J. Wu, "Energy Efficiency in Cloud Computing: Techniques and Challenges," *IEEE Trans. Cloud Comput.*, vol. 6, no. 3, pp. 695–708, 2018. DOI: 10.1109/TCC.2017.2779125.

[23] H. Fang and Y. Zhang, "Energy-Efficient Cloud Data Centers: A Survey on Recent Advances and Future Directions," *Comput. Electr. Eng.*, vol. 82, p. 106564, 2020. DOI: 10.1016/j.compeleceng.2019.106564.

[24] J. Gomez and F. Garcia, "Sustainable Data Center Design and Optimization," *J. Comput. Netw. Commun.*, vol. 2017, pp. 1–16, 2017. DOI: 10.1155/2017/5308097.

[25] C. Li and L. Yang, "Green Cloud Computing and Sustainable Data Centers: A Review," *IEEE Access*, vol. 4, pp. 9313–9330, 2016. DOI: 10.1109/ACCESS.2016.2615138.

[26] X. Xia and W. Zhang, "Energy-Efficient Resource Management in Cloud Computing: Techniques and Strategies," *ACM Comput. Surv.*, vol. 54, no. 4, pp. 1–35, 2021. DOI: 10.1145/3447797.

[27] N. Jones, "How to stop data centres from gobbling up the world's electricity," *Nature*, vol. 561, no. 7722, pp. 163–166, 2018.

[28] J. G. Koomey, *Growth in Data Center Electricity Use 2005 to 2010*. Analytics Press, pp. 1–9, 2011.

[29] A. Shehabi et al., *United States Data Center Energy Usage Report*. Lawrence Berkeley Nat. Lab., pp. 1–87, 2016.

[30] C. Belady, A. Rawson, J. Pfleuger, and T. Cader, "Green Grid Data Center Power Efficiency Metrics: PUE and DCiE," *The Green Grid Consortium*, vol. 6, no. 2, pp. 1–9, 2008.

[31] T. Levin and V. M. Thomas, "Evaluating the energy savings potential of large-scale data center consolidations," *J. Energy Eng.*, vol. 142, no. 2, pp. 1–8, 2016.

[32] A. Qureshi, R. Weber, H. Balakrishnan, J. Guttag, and B. Maggs, "Cutting the electric bill for internet-scale systems," *ACM SIGCOMM Comput. Commun. Rev.*, vol. 39, no. 4, pp. 123–134, 2009.

[33] C. D. Patel, R. K. Sharma, C. E. Bash, and S. Graupner, "Energy-aware distributed computing with adaptive hierarchical load balancing," *Hewlett-Packard Labs. Tech. Rep.*, pp. 1–11, 2002.

[34] R. K. Rongon, M. Akhtaruzzaman, K. Das and S. T. Sharif, "Analyzing Risk Level from RoI and VoI for Robust Data Privacy using Fuzzy Inference System," 2025 4th International Conference on Robotics, Electrical and Signal Processing Techniques (ICREST), Dhaka, Bangladesh, 2025, pp. 421-425, doi:10.1109/ICREST63960.2025.10914468.

[35] R. K. Rongon, and K. Das, "An energy-efficient cloud data center architectural model for the national data center of Bangladesh," J. Discrete Math. Sci. Cryptogr., vol. 27, no. 6, pp. 1905–1917, 2024, doi: 10.47974/JDMSC-1916.

[36] R. K. Rongon, K. Das, "Obstacle to implement sustainable computing in Bangladesh and approach to being sustainable through green cloud computing methodology," International Conference on Sustainable Computing (SUSCOM-2022), Dhaka, Bangladesh, Volume: 2.